\documentclass[10pt,twocolumn,a4paper,letter]{IEEEtran}
\usepackage{graphicx}
\usepackage{caption,color}
\usepackage{refstyle}
\usepackage{amsmath}
\usepackage{amsfonts}
\usepackage{subfigure}
\usepackage{algorithm}
\usepackage{amssymb}
\usepackage{cite}
\usepackage{algorithmic}
\usepackage{epstopdf}
\usepackage{cite}
\usepackage{booktabs}
\usepackage{slashbox}
\usepackage{amsthm}
\usepackage{subfigure}

\newcommand\numberthis{\addtocounter{equation}{1}\tag{\theequation}}

\DeclareMathOperator*{\argmax}{arg\,max}
\ifCLASSINFOpdf
\else
\fi
\begin{document}
\linespread{1.0}
\title{Compute-and-Forward in Cell-Free Massive MIMO: Great Performance with Low Backhaul Load}
\author{Qinhui Huang and Alister Burr\\University of York, U.K.\\\{qh529,alister.burr\}@york.ac.uk
\thanks{The work described in this paper was supported in part by UK EPSRC under grant EP/K040006.}}

\maketitle

\thispagestyle{empty}

\begin{abstract}
In this paper, we consider the uplink of cell-free massive MIMO systems, where a large number of distributed single antenna access points (APs) serve a much smaller number of users simultaneously via limited backhaul. For the first time, we investigate the performance of compute-and-forward (C\&F) in such an ultra dense network with a realistic channel model (including fading, pathloss and shadowing). By utilising the characteristic of pathloss, a low complexity coefficient selection algorithm for C\&F is proposed. We also give a greedy AP selection method for message recovery. Additionally, we compare the performance of C\&F to some other promising linear strategies for distributed massive MIMO, such as small cells (SC) and maximum ratio combining (MRC). Numerical results reveal that C\&F not only reduces the backhaul load, but also significantly increases the system throughput for the symmetric scenario.  
\end{abstract}

\section{introduction}
Massive MIMO is a promising technique to meet the capacity density requirement in 5G wireless. By increasing the ratio of BS antennas to users, wireless networks tend to a quasi orthogonal, `interference free' state\cite{reference13}\cite{reference8}. Recently, distributed massive MIMO has attracted a lot interest. Compared to the collocated massive MIMO, the distributed version brings the APs much closer to the the `cell edge' users, which leads to a uniformly good service for all users. A traditional way to perform such distributed massive MIMO is by means of a small-cell\cite{reference14} deployment, where users benefit from selection combining of denser APs. Recently, the authors in \cite{reference9}\cite{reference10} proposed a `cell-free' model, where all APs serve all users simultaneously. Simple maximum ratio combining (MRC) is employed for the uplink scenario at all antenna sites. They showed that cell free massive MIMO gives further improvement for `cell edge' users compared to the small-cell scheme.\par 
However, the assumption of infinite backhaul in \cite{reference9}\cite{reference10} is not feasible in practice. Reducing the backhaul load is usually the main challenge in any distributed antenna systems. Compute-and-forward (C\&F)\cite{reference2} is an efficient approach for backhaul reduction. It employs structured lattice codes for physical layer network coding. Each AP infers and forwards an integer combination of the transmitted symbols of all users; hence cardinality expansion is avoided. The performance of C\&F greatly depends on the coefficient selection of the integer equations. This comprises two aspects: on the one hand, each AP needs to selecte its local best equation corresponds to the highest computation rate. Many efficient algorithms \cite{reference5, reference12, reference15, reference16} have been developed during the past two years. On the other hand, the integer equations provided by all APs should form a full rank matrix, in order to recover messages of all users. Thanks to the large ratio of AP/user antennas in massive MIMO, it is then likely that randomly-chosen coefficients result in a full rank matrix, and hence this is not a serious problem.\par 
As mentioned above, the primary advantage of cell-free massive MIMO is to provide uniformly good service for all users, hence we focus on the symmetric scenario where all users transmit with a common rate. We consider a network that contains a large number of users, and take the pathloss and shadowing into account. To the best of our knowledge, C\&F has not previously been studied in such a scenario. The main contributions, of this paper are as follows:
\begin{itemize}
\item Exploiting the properties of pathloss, we propose a novel coefficient selection algorithm to further reduce the complexity.
\item We propose a greedy AP selection algorithm for message recovery at the central hub. 
\item We study the performance of C\&F in cell-free massive MIMO systems from three perspectives: 1) the probability of rank deficiency. 2) the outage probability for a given rate. 3) the achievable throughput.
\item We provide a comprehensive comparison between C\&F, small cells and MRC. We show that C\&F achieves the best performance among the three schemes. Their respective complexities and required backhaul are also discussed.         
\end{itemize} 
\par
The rest of this paper is organised as follows. We briefly review the cell-free model and C\&F strategy in Section II, and introduce the coefficient selection algorithm and the AP selection method in Section III. Numerical results and discussions are given in Section IV. We conclude the paper in Section V.  \par 

Unless noted, we use plain letters, boldface lowercase letters and boldface uppercase letters to denote scalars, vectors and matrices respectively. All vectors are column vectors. The set of real numbers and complex numbers are represented by $\mathbb{R}$ and $\mathbb{C}$ respectively. Transpose, Hermitian transpose, and the round operation are represented by $[\cdot]^T$, $[\cdot]^H$ and $\lfloor\cdot\rceil$ respectively. We use $\|\cdot\|$ to denote the Euclidean
norm.          
\section{Preliminaries}
\subsection{System Model}
We consider a system model which is similar to the cell free massive MIMO in \cite{reference9}. There are $M$ APs and $L$  ($M>L$) users randomly deployed in a large square area, and all users share the same time-frequency resource. The main difference here is that the APs are connected via limited (rather than infinite) backhaul (or fronthaul) to a hub.\par
\begin{center}
\begin{figure}[h]
\centerline{\includegraphics[width=0.95\linewidth]{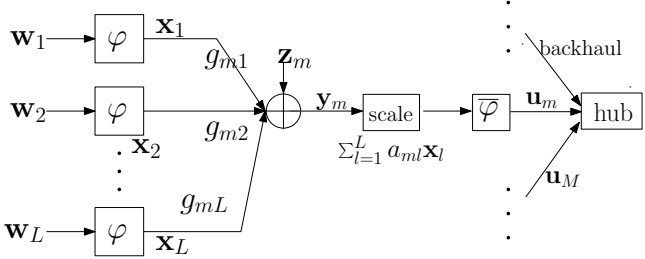}}
\caption{C\&F in cell free massive MIMO}
\label{fig:buffer1}
\end{figure}
\end{center}
\par 
As shown in Fig.1, the length-$k$ data of the $l$th user $\mathbf{w}_{l}\in{\mathbb{F}_{p}^{k}}$ is encoded into an length-$n$ codeword $\mathbf{x}_{l}\in{(\mathbb{Z}[i]/{\pi}\mathbb{Z}[i])^n}$. \footnote{$\mathbb{F}_{p}$ denotes the finite field. $\mathbb{Z}[i]$ denotes the Gaussian integers whose real and imaginary parts are both integers. The ring quotient $\mathbb{Z}[i]/{\pi}{\mathbb{Z}[i]}$ is isomorphic to $\mathbb{F}_p$. See \cite{reference17, reference7} for more details.} The codebook is denoted by a ring quotient of fine lattice $\Lambda$ and coarse lattice $\Lambda'$, written as $\Lambda/{\Lambda'}$.\footnote{Note that $\mathbb{Z}[i]/{\pi}\mathbb{Z}[i]$ is defined on symbol, whereas $\Lambda/{\Lambda'}$ is defined on codeword, their respective cardinalities are $p$ and $p^{k}$.} We assume the power constraint of each codeword is $\mathbb{E}[\|{\mathbf{x}_{l}}\|]^2\leq{nP}$. The received length-$n$ vector of the $m$th AP can be expressed as
\begin{equation}
\mathbf{y}_m=\mathbf{X}^{T}\mathbf{g}_m+\mathbf{z}_{m},\mathbf{y}_m\in{\mathbb{C}^{n}},
\end{equation}
where $\mathbf{X}=[\mathbf{x}_{1},\mathbf{x}_2,\cdots,\mathbf{x}_{L}]^{T}$ denotes the signal matrix, and the thermal noise $\mathbf{z}_m\sim{\mathcal{CN}(0,\sigma^2{\mathbf{I}_{n}})}$ is a circularly symmetrical complex Gaussian random vector. We assume there is no pilot contamination, which means perfect channel estimation is available at APs. The channel coefficient vector is denoted by $\mathbf{g}_{m}=[g_{m1},g_{m2},\cdots,g_{mL}]^T$, where $g_{ml}$ represents the channel link between the $l$th user and the $m$th AP, given as
\begin{equation}
g_{ml}=\sqrt{\mathbf{PL}_{ml}\mathbf{S}_{ml}}h_{ml},g_{ml}\in{\mathbb{C}},
\end{equation}
where $\mathbf{PL}_{ml}$, $\mathbf{S}_{ml}$ and $h_{ml}$ denotes the pathloss, shadowing and small-scale fading respectively. \par
Each AP attempts to use an integer linear combination of the codewords to represent the scaled received signal, expressed as $\mathcal{Q}_{\Lambda}(\alpha_{m}\mathbf{y}_{m})=\mathbf{a}_{m}^{T}\mathbf{X}$. Here $\mathcal{Q}_{\Lambda}$ quantises $\alpha_{m}\mathbf{y}_m$ to its closest fine lattice point in $\Lambda$. The quantisation error contributes to the effective noise of the C\&F scheme, whose variance can be expressed as
\begin{equation}
\sigma_{m}^2=\|\alpha_{m}\mathbf{g}_{m}-\mathbf{a}_{m}\|^2P+\|\alpha_{m}\|^2{\sigma}^2.
\end{equation}
For a given integer coefficient vector $\mathbf{a}_m=[a_{m1},a_{m2},\cdots,a_{mL}]^T$, there exists an optimal scaling factor $\alpha_{\mathrm{opt}}=\frac{\mathrm{SNR}\mathbf{g}_{m}^{H}\mathbf{a}_{m}}{1+\mathrm{SNR}\|\mathbf{g}_{m}\|^2}$ to minimise the effective noise, with $\mathrm{SNR}=P/{\sigma^2}$. Hence the achievable computation rate is given by \cite{reference2}
\begin{equation}
\mathcal{R}_{m,\mathrm{C\&F}}(\mathbf{g}_{m},\mathbf{a}_m)=\mathrm{log}_{\mathrm{2}}^{\dagger}\Big(\displaystyle{\frac{1}{\mathbf{a}_{m}^{H}\mathbf{M}\mathbf{a}_{m}}}\Big),
\end{equation}
where $\mathbf{M}=\mathrm{I}_{L}-\displaystyle{\frac{\mathrm{SNR}}{\mathrm{SNR}\|\mathbf{g}_{m}\|^2+1}}\mathbf{g}_{m}\mathbf{g}_{m}^{H}$, and $\mathbf{I}_{L}$ is an $L\times{L}$ identity matrix. The target of each AP is to find its local best integer vector to maximise the computation rate, written as
\begin{equation}
\mathbf{a}_{m,\mathrm{opt}}=\displaystyle{\argmax_{\mathbf{a}_m\in{{\mathbb{Z}[i]}^{L}\setminus{\{\mathbf{0}\}}}}\mathcal{R}(\mathbf{g}_m,\mathbf{a}_{m})}.
\end{equation}
\par
Finally, a lattice decoder $\overline{\varphi}$ is used to decode the linear combination of the codewords to the linear combination of the original data, expressed as
\begin{equation}
\mathbf{u}_{m}=\overline{\varphi}(\sum_{l=1}^{L}{a_{ml}}\mathbf{x}_{l})=\sum_{l=1}^{L}{q_{ml}}\mathbf{w}_{l},
\end{equation} 
with $q_{ml}\in\mathbb{F}_{p}$ and $\mathbf{u}_{m}\in\mathbb{F}_{p}^{k}$, the cardinality remains the same as the original data.
\subsection{Benchmarks}
\subsubsection{Maximum Ratio Combining (MRC)}
In the original paper of cell free massive MIMO \cite{reference10}, the received signal at the $m$th AP is multiplied by the conjugate transpose of the channel vector $\mathbf{g}_{m}$; then the precise signal matrix of $\mathbf{y}_{m,\mathrm{MRC}}=\mathbf{y}_{m}\mathbf{g}_{m}^{H}$ is forwarded to the hub via infinite backhaul. The hub combines the received signal of all APs, hence the achievable rate of the $l$th user can be expressed as
\begin{equation}
\mathcal{R}_{l,\mathrm{MRC}}=\mathrm{log}_{2}\Big(1+\displaystyle{\frac{\mathrm{SNR}\|\mathbf{g}_{l}\mathbf{g}_{l}^{H}\|^2}{\|\mathbf{g}_{l}^{H}\|^2+\mathrm{SNR}\displaystyle{\sum_{i\neq\l}\|\mathbf{g}_{i}\mathbf{g}_{l}^{H}\|^2}}}\Big),
\end{equation}
where $\mathbf{g}_{l}=[g_{1l},g_{2l},\cdots,g_{Ml}]$ is the row channel vector corresponding to the $l$th user. Since we concentrate on the symmetric scenario, the system throughput per user is determined by the worst user, denoted as
\begin{equation}
\mathcal{R}_{\mathrm{MRC}}=\min_{l=1,\cdots,L}\mathcal{R}_{l,\mathrm{MRC}}.
\end{equation}
\subsubsection{Small Cells}
In small cell deployment, we assume each user selects one AP only among all APs based on the strength of the channel link $\mathbf{g}_{ml}$. \footnote{Since in C\&F scheme, the coefficient selection is performed during the coherence time of small scale fading, we assume the AP allocation in small cell is also performed during the small scaling fading coherence time, in order to provide a fair comparison.} The APs allocation is performed user by user with random priorities. Assuming $m_{l}$ is the index of the AP allocated to the $l$th user, its achievable rate can be expressed as
\begin{equation}
\mathcal{R}_{l,\mathrm{SC}}=\mathrm{log}_{2}\Big(1+\displaystyle{\frac{\mathrm{SNR}\|g_{m_{l}l}\|^2}{1+\mathrm{SNR}\displaystyle{\sum_{l'\neq{l}}}\|g_{m_{l}l'}\|^2}}\Big).
\end{equation}      
Again, the symmetric rate depends on the worst user.
\section{C\&F in cell free massive MIMO}

\subsection{Why is Compute and Forward Good for Cell-free?}
In this section, we provide an intuitive analysis of C\&F in the cell free massive MIMO systems, in terms of backhaul, complexity and throughput.
\subsubsection{backhaul} Clearly, this is the primary advantage of C\&F compared to any other linear processing schemes. The cardinality required at each AP is the same as the cardinality of the user data, this is theoretically the minimum backhaul required to achieve lossless transmission.
\subsubsection{Complexity} Compared to small cell and MRC, the extra complexity of C\&F arises from the coefficient selection. In a quasi static case, this additional complexity is negligible compared to channel coding and decoding. 
\subsubsection{Throughput} In small cells, users are served by APs in a one-to-one manner. In C\&F, each equation provided by one AP might involve one or more users, and each user might take part in multiple equations from many APs. Hence we conclude that small cell is a special case of C\&F in which only one user is involved in each equation, and we therefore expect C\&F to achieve higher throughput than small cells.\par
MRC enables each user to be served by all APs. Unlike the collocated massive MIMO, the channel strength for a specific user varies at different APs. MRC eliminates inter-user interference only asymptotically, as ratio of antennas to users tend to infinity. In contrast, provided a full rank matrix is formed, C\&F allows all users to be recovered, analogously to zero-forcing, but without noise enhancement. Hence C\&F may also outperform MRC.     

\subsection{Coefficient Selection}
Previously we have stated that each AP aims to find an integer vector to maximise the computation rate locally, expressed as (5). The latest research \cite{reference12, reference15, reference16} show that it is more convenient to optimise the scaling factor $\alpha_{m}$ directly, hence the optimisation problem of (5) is translated to the following expression
\begin{align*}
\alpha_{m,\mathrm{opt}}&=\argmax_{\alpha_{m}\in{\mathbb{C}}\setminus{\{0\}}}\mathcal{R}(\mathbf{g}_{m},\alpha_{m})\\&=\argmax_{\alpha_{m}\in{\mathbb{C}}\setminus{\{0\}}}\mathrm{log}_{2}^{\dagger}\Big(\displaystyle{\frac{P}{\|\alpha_{m}\|^2{\sigma}^2+\|\alpha_{m}\mathbf{g}_{m}-\lfloor{\alpha_{m}\mathbf{g}_{m}\rceil}\|^2{P}}}\Big)\numberthis,
\end{align*} 
where $\lfloor\cdot\rceil$ denotes the quantisation to its closest Gaussian integer, and $\lfloor\alpha_{m}\mathbf{g}_{m}\rceil$ is the corresponding integer vector $\mathbf{a}_{m}$ of $\alpha_{m}$. The time complexity for such an optimisation problem is $\mathcal{O}(L\mathrm{log}(L)\mathrm{SNR})$. Now we introduce a novel method to reduce the complexity.\par 
Since the computation rate has to be an non-negative value, hence we can easily obtain the upper bound of $\alpha_{m}$ from (10), given by
\begin{equation}
\alpha_{m,\mathrm{ub}}=\sqrt{P/{\sigma^2}}=\sqrt{\mathrm{SNR}}.
\end{equation}
For the users which are located far away from the $m$th AP, the amplitudes of their corresponding channel $g_{ml}$ are usually very small. Even multiplied by the upper bound of $\alpha_m$, their corresponding integer coefficients are still zero, expressed as
\begin{equation}
a_{ml}=\lfloor\sqrt{\mathrm{SNR}}{g_{ml}}\rceil=0.
\end{equation}
This means these users are not able to contribute to the linear equation of the $m$th AP, no matter what $\alpha_m$ is selected. Hence these users can be simply treated as interference, which adds to the thermal noise, and the number of effective users becomes
\begin{equation}
L_{\mathrm{eff}}=|\Phi|,\Phi=\{l:\lfloor\sqrt{\mathrm{SNR}}{g_{ml}}\rceil\neq{0}\},
\end{equation}
where $|\Phi|$ denotes the cardinality of set $\Phi$. The effective SNR becomes
\begin{equation}
\mathrm{SNR}_{\mathrm{eff}}=\displaystyle{\frac{P}{\sigma^2+\sum_{l\notin{\Phi}}^{L}\|g_{ml}\|^2}},
 \end{equation}
 hence the time complexity becomes $\mathcal{O}(L_{\mathrm{eff}}\mathrm{log}(L_{\mathrm{eff}})\mathrm{SNR_{eff}})$, clearly less than $\mathcal{O}(L\mathrm{log}(L)\mathrm{SNR})$. The proportion of the reduced complexity depends on the density of users: the details will be presented in a subsequent paper.
\subsection{Data Recovery at Hub}
The $M$ linear equations received by the hub form an $M\times{L}$ matrix, written as $\mathbf{A}=[\mathbf{a}_1,\mathbf{a}_2,\cdots,\mathbf{a}_M]^T$. \footnote{The transfer matrix received by hub is actually $\mathbf{Q}$ which is made up of $q_{ml}$ in (6). However, the assessment of $\mathbf{Q}$ relies on a specific finite field $\mathbb{F}_{q}$. Hence the integer matrix $\mathbf{A}$ is commonly used instead of $\mathbf{Q}$ for the performance evaluation in a general case. See \cite{reference18} for details.} We define 
\begin{equation}
M_{\mathrm{rank}}\triangleq\mathrm{Rank}(\mathbf{A}) 
\end{equation} 
as the maximum number of users whose data can be recovered. Clearly, $1\leq{M_{\mathrm{rank}}}\leq{L}$, and all users can be recovered iff $M_{\mathrm{rank}}=L$. In the traditional multiuser MIMO systems ($M=L$), the data recovery of C\&F is a major challenge due to the high probability of rank deficiency. We expect however that the extra APs in massive MIMO can ensure a much higher probability that $M_{\mathrm{rank}}$ is equal or at least close to $L$. \par

\begin{algorithm}
\caption{Greedy AP selection for message recovery}
\begin{algorithmic}[1]
\REQUIRE $\mathbf{A}$, $\mathcal{R}_{m,\mathrm{C\&F}}$ with $m\in\{1,2,\cdots,M\}$ 
\ENSURE selected APs (equations): $\mathbf{A}_{\mathrm{sub,opt}}$\\*
Initialisation :
$i=1$, $discard=0$, $M_{\mathrm{rank}}=\mathrm{Rank}(\mathbf{A})$

\STATE sort $\mathcal{R}_{m,\mathrm{C\&F}}$ in ascending order, with indices of sorted $\mathcal{R}$ in an $1\times{M}$ vector $\Theta$. \\
$\mathbf{A}_{\mathrm{sub,opt}}=[\mathbf{a}_{\Theta(1)},\mathbf{a}_{\Theta(2)},\cdots,\mathbf{a}_{\Theta(M)}]^{T}$

\IF {$\mathrm{Rank}(\mathbf{A}_{\mathrm{sub,opt}}\setminus{\mathbf{a}_{\Theta(i)}})=\mathrm{Rank}(\mathbf{A}_{\mathrm{sub,opt}})$}
\STATE set $\mathbf{A}_{\mathrm{sub,opt}}
\gets{\mathbf{A}_{\mathrm{sub,opt}}\setminus\mathbf{a}_{\Theta{(i)}}}$.\\$discard\gets{discard+1}$
\ENDIF
\STATE set $i\gets{i+1}$, terminate if $discard=M-M_{\mathrm{rank}}$,\\
otherwise go to line 2.
\STATE Return $\mathbf{A}_{\mathrm{sub,opt}}$

\end{algorithmic}
\end{algorithm}

Let $\mathbf{A}_{\mathrm{sub}}$ to denote a submatrix of $\mathbf{A}$ which is formed by taking rows of $\mathbf{A}$ with indices in $\mathcal{S}\subset\{1,2,\cdots,M\}$. $\mathbf{A}_{\mathrm{sub}}$ has to meet the conditions of $\mathrm{Rank}(\mathbf{A}_{\mathrm{sub}})=|\mathcal{S}|=M_{\mathrm{rank}}$. The optimal strategy for message recovery is to find such an $\mathbf{A}_{\mathrm{sub}}$ to maximise the corresponding computation rate of the worst row (worst equation) of $\mathbf{A}_{\mathrm{sub}}$. This max-min optimisation problem can be expressed as
\begin{equation}
\mathbf{A}_{\mathrm{sub,opt}}=\argmax_{\mathbf{A}_{\mathrm{sub}}: \mathrm{Rank}(\mathbf{A}_{\mathrm{sub}})=|\mathcal{S}|=M_{\mathrm{rank}}}\min_{m\in{\mathcal{S}}}\mathcal{R}_{m,\mathrm{C\&F}}.
\end{equation}  \par 
 
We propose a simple greedy algorithm to acquire the optimal solution. We sort the rows of $\mathbf{A}$ in ascending order of their computation rates, and check them one by one. For each row, if its absence would not change the rank, then it can be discarded (clearly, it can be replaced by another equation with higher rate). This procedure terminates when the remaining rows of $\mathbf{A}$ meet the rank requirement. The greedy AP selection method is summarised in Algorithm 1. \par

Note that we have two options to utilise the AP selection:
\begin{itemize}
\item All APs transmit their messages via $R_{0}$, and the hub collects information from all $M$ APs. Hence the AP selection is utilised only for the message recovery. The total backhaul load is $MR_{0}$ 
\item At the beginning of each coherence time, all APs send their local best coefficient vectors and corresponding rates to the hub, and then hub selects the APs and gives feedback (Acknowledgement) to the selected APs, hence only $M_{\mathrm{rank}}$ APs are active during each transmission. The total backhaul in this case is $M_{\mathrm{rank}}R_{0}$.
\end{itemize}  
Since the system throughput is determined only by which APs are selected, the two strategies above have the same performance. The former minimises the latency, while the latter minimises backhaul.
\section{Numerical results and discussions }
In this section, we provide numerical results of C\&F performance on cell free massive MIMO systems, and compare it with MRC and small cells. 
\subsection{Simulation Setup}
We adopt the parameter settings in \cite{reference9} as the basis to establish our simulation model. In all examples, the users and APs are randomly uniformly distributed in a square of $1\times{1}\mathrm{km}$. Due to the wide range of the square, we assume independent small scaling fading $h_{ml}\sim{\mathcal{CN}(0,1)}$ and uncorrelated shadowing $\mathbf{S}_{ml}$ for different AP-user pairs. The shadowing is denoted as
\begin{equation}
\mathbf{S}_{ml}=10^{\frac{\sigma_{\mathrm{sh}}{z_{ml}}}{10}},
\end{equation}
where the standard deviation $\sigma_{\mathrm{sh}}$ is set to 8dB, and $z_{ml}\sim{\mathcal{N}(0,1)}$.
\par
The pathloss $\mathbf{PL}_{ml}$ is simulated by a three-slope model. The exponent relies on the distance between the $l$th user and the $m$th AP, denoted as $d_{ml}$. It is equal to 0, 2 and 3.5, for $d_{ml}\leq{10}\mathrm{m}$, $10\mathrm{m}<{d_{ml}}\leq{50\mathrm{m}}$ and $50\mathrm{m}<d_{ml}$ respectively. The Hata-COST231 model is employed to characterise the propagation assuming that the carrier frequency is 1.9GHz. The heights of APs and users are 15m and 1.65m respectively. \par 
The other parameters are chosen as the most commonly used values. The system bandwidth is 20MHz, and the transmit power and the thermal noise density are set to 200mW and -174dBm/Hz respectively. Equal power allocation is assumed in all examples.\par
\subsection{The Probability of Rank Deficiency }    In this section, we investigate the probability of rank-deficiency of C\&F with different AP/user ratios. Fig. 2 shows the cumulative distribution of  $M_{\mathrm{rank}}$ over 200 channel realisations. As discussed previously, the traditional MU-MIMO ($M=L$) severely suffers from rank-deficiency, represented as the solid thin line. The range of $M_{\mathrm{Rank}}$ is between 21 to 32, this means none of the channel realisation corresponds to a full-rank matrix, and the maximum number of decodable users is only 32. When the number of APs is increased to 100, the probability of full-rank becomes 22.5\%, the at least 35 users are supported by the integer matrix. When we further increase $M$ to 200, the full-rank probability becomes 96\%, and the minimum number of users get involved is 39. Therefore, we conclude the rank-deficiency problem is trivial in a massive MIMO scenario.         
\begin{center}
\begin{figure}[h]
\centerline{\includegraphics[width=1.05\linewidth]{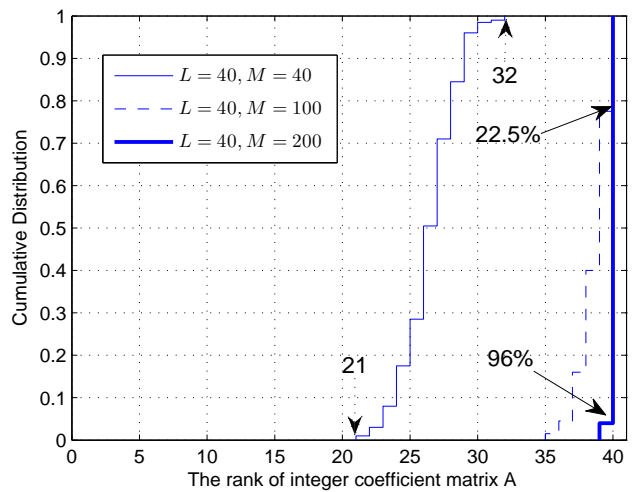}}
\caption{The CDF of $M_{\mathrm{rank}}$ with different AP/user ratios}
\label{fig:buffer1}
\end{figure}
\end{center}

\subsection{Outage Probability}
We now compare the performance of C\&F, MRC and small cells. Fig. 3 illustrates their achievable rates under one example channel realisation. There are 40 users and 100 APs in this example. The black circles calculated by (7) and red squares calculated by (9) represent the rates of 40 users for MRC and small-cell respectively. The blue crosses denote the corresponding computation rates of equations in $\mathbf{A}_{\mathrm{sub,opt}}$ (if $M_{\mathrm{rank}}<40$, use zero to denote the rest $40-M_{\mathrm{rank}}$ computation rates). All rates are sorted in ascending order, and the y-axis denotes the index of users (or equations for C\&F).\par We can see that for both C\&F and small cell schemes, the corresponding rates of the top 4 users (or equations) are exactly the same (shown in the right top corner of Fig. 3). This is due to the fact that when an user is located very close to a specific AP, the equation provided by that AP is very likely to contain that user only, hence the C\&F is equivalent to small cells. For other APs, C\&F provides higher rate equations compared to the `single user access' in a small cell. Thus small-cell is a special case of C\&F as discussed in section III.$A$. It is also observed that the C\&F scheme gives the best performance for `cell edge' users. Assuming that all users transmit with rate $R_0=0.5$, the number of outage users ($R<R_{0}$, corresponding to the points located on the left of the black dashed line) for C\&F, MRC and small cells are 1, 9 and 10 respectively.           
\begin{center}
\begin{figure}[h]
\centerline{\includegraphics[width=1.05\linewidth]{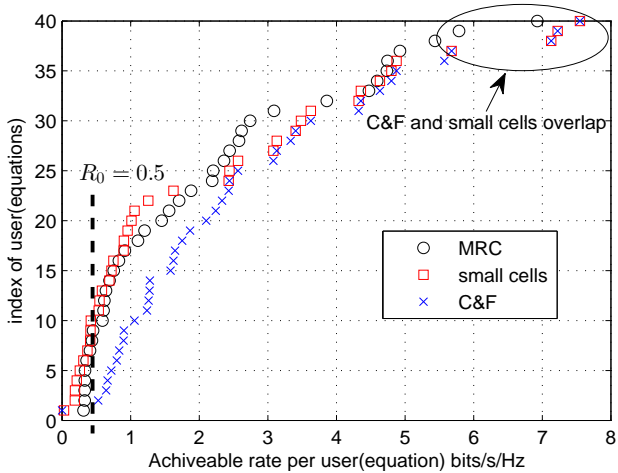}}
\caption{Rate comparison for one channel realisation }
\label{fig:buffer1}
\end{figure}
\end{center}

\begin{figure}[htp]
  \begin{center}
    \subfigure[$L=40$, $M=100$]{\label{fig:edge-a}\includegraphics[scale=0.70]{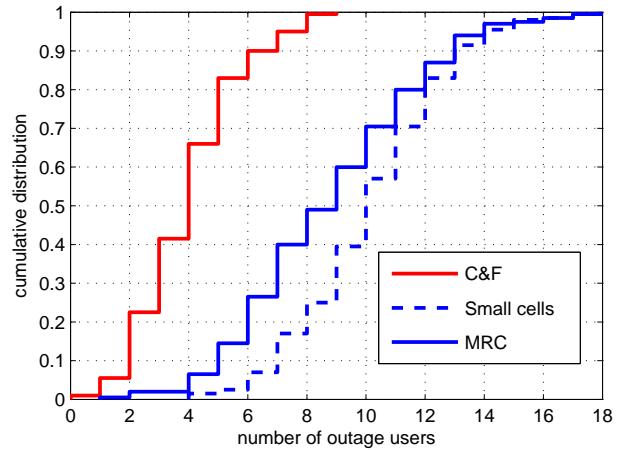}}
    \subfigure[$L=40$, $M=200$]{\label{fig:edge-b}\includegraphics[scale=0.70]{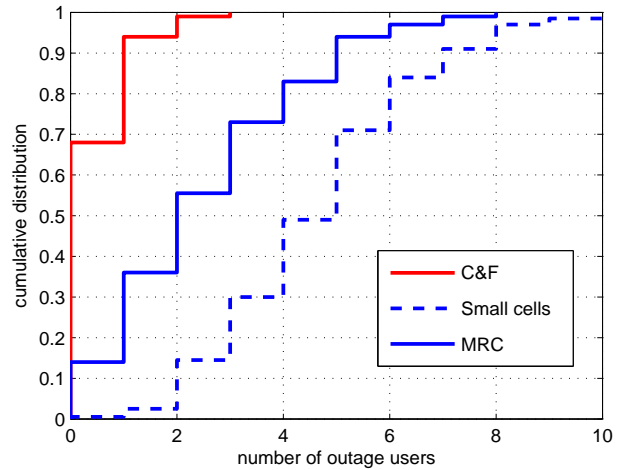}} 
  \end{center}
  \caption{The CDF of number of outage user $N_{\mathrm{outage}}$}
  \label{fig:edge}
\end{figure}

Fig. 3 has shown C\&F has the smallest number of outage users (denoted as $N_{\mathrm{outage}}$) for a specific channel. Now we evaluate the outage performance in a more general scenario. Fig. 4 illustrates the cumulative distribution of $N_{\mathrm{outage}}$ over 200 channel realisations. Again we set $R_{0}=0.5$ as the target rate. As expected, C\&F outperforms the other two schemes in both the $M=100$ and $M=200$ scenarios (We include among the outage users in C\&F also the undecodable users due to rank-deficiency). Particularly for the case of $M=200$, about 70\% of channel realisations achieve outage-free transmission employing C\&F, while the other two schemes still suffer nearly 10 user outages for some channel realisations.    
\subsection{Outage Rate}
In the previous section, we investigated the outage performance with a given target rate. We define the outage probability to be the expected value of $N_{\mathrm{outage}}/L$, denoted as
\begin{equation}
\rho_{\mathrm{outage}}(R_{0})\triangleq\mathrm{Pr_{outage}}(R<R_{0})=\mathbb{E}[\frac{N_{\mathrm{outage}}}{L}].
\end{equation}
In this section, we will characterise the performance by its outage rate for a given target outage probability, defined as
\begin{equation}
R_{\mathrm{outage}}(\rho)\triangleq\mathrm{sup}\{R:\rho_{\mathrm{outage}}(R)\leq{\rho}\},
\end{equation}
where sup stands for `supremum'. Recall that we proposed an alternative scheme in III.$C$. When some feedback is available, the AP selection can be done before each transmission. This means a subset of APs and users can be scheduled as active for each transmission. The outage rate is exactly the throughput per user for such an alternative C\&F scheme. \par
For example, if we allow an outage probability 1/8, then we only need to schedule $40\times(1-1/8)=35$ users and APs for each transmission. The corresponding outage rate (throughput) is determined by the 6th worst user (for MRC and small cells) or equations (for C\&F). Again, we treat the equations corresponding to insufficient rank as 0 rate equations.\par 
As shown in Fig. 5 (a), when $M=100$ and $\rho_{\mathrm{outage}}=1/8$, for the 95\%-likely channel realisations, the throughput of C\&F is 2 and 3 times better than MRC and small cells respectively. Fig. 5 (b) reveals a similar advantage for C\&F, furthermore in an 100\%-likely manner. Fig. 5 (c) illustrates that by increasing the AP/user ratio, C\&F retains a significant advantage even for a much lower outage probability ($\rho_{\mathrm{outage}}=1/40$). It also means more users can be scheduled for each transmission.   
\begin{figure}[htp]
  \begin{center}
    \subfigure[$L=40$, $M=100$, $\rho_{\mathrm{outage}}=1/8$]{\label{fig:edge-a}\includegraphics[scale=0.58]{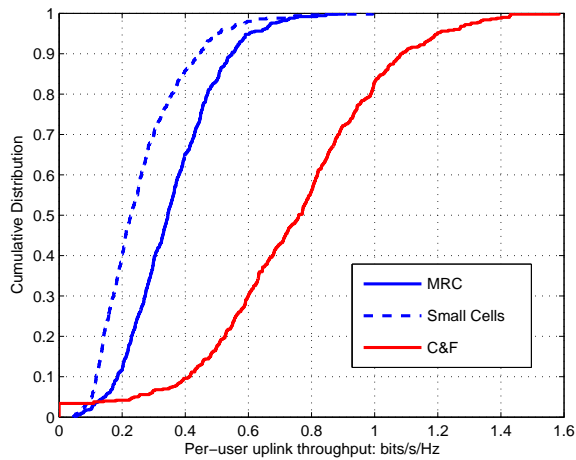}}
    \subfigure[$L=40$, $M=200$, $\rho_{\mathrm{outage}}=1/8$]{\label{fig:edge-b}\includegraphics[scale=0.58]{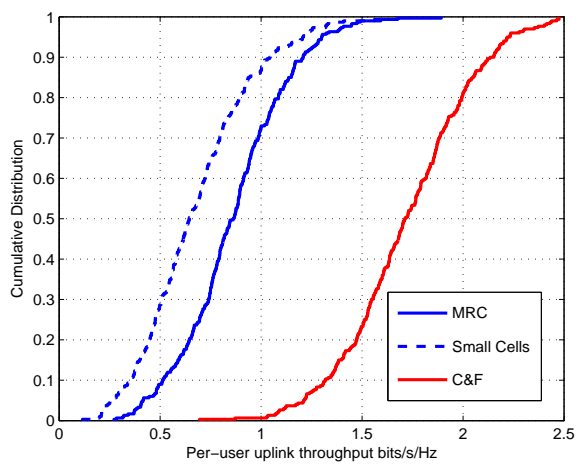}} \\
    \subfigure[$L=40$, $M=200$, $\rho_{\mathrm{outage}}=1/40$]{\label{fig:edge-a}\includegraphics[scale=0.58]{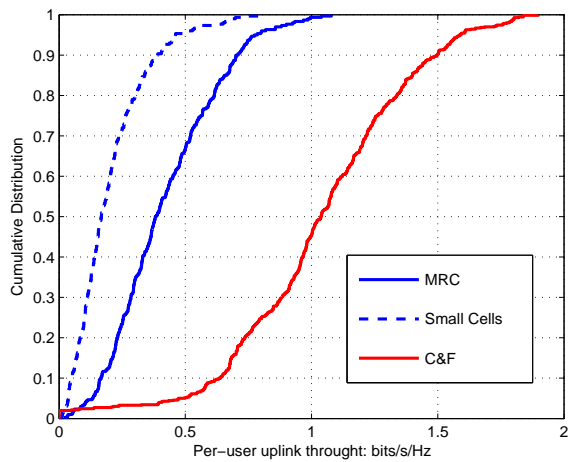}}
  \end{center}
  \caption{The CDF of per-user throughput (outage rate)}
  \label{fig:edge}
\end{figure}

\section{concluding remarks}
In this paper, we applied C\&F scheme to the cell free massive MIMO systems to reduce the backhaul load, and analysed its benefits in such systems. A novel low complexity algorithm for coefficient selection is proposed. We also presented a simple greedy method for AP selection.  
Numerical results have shown that C\&F outperforms MRC and small cells, in terms of both outage probability and system throughput. 

\bibliographystyle{IEEEtran}
\bibliography{ICC16}

\end{document}